\documentclass[aps,prl,twocolumn,noshowpacs,groupedaddress]{revtex4}  
\usepackage{graphicx}  
\usepackage{dcolumn}   
\usepackage{bm}        
\usepackage{amssymb}   
\usepackage{multirow}

\begin{document}

\title{A microfabricated surface-electrode ion trap for scalable quantum information processing}

\author{S.~Seidelin}
\email{seidelin@boulder.nist.gov}
\author{J.~Chiaverini}
\altaffiliation[Present Address: ]{Los Alamos National Laboratory}
\author{R.~Reichle}
\author{J.~J.~Bollinger}
\author{D.~Leibfried}
\author{J.~Britton}
\author{J.~H.~Wesenberg}
\author{R.~B.~Blakestad}
\author{R.~J.~Epstein}
\author{D.~B.~Hume}
\author{J.~D.~Jost}
\author{C.~Langer}
\author{R.~Ozeri}
\author{N.~Shiga}
\author{D.~J.~Wineland}
\affiliation{NIST, Time and Frequency Division, Boulder, CO}
\date{\today}

\begin{abstract}
We demonstrate confinement of individual atomic ions in a
radio-frequency Paul trap with a novel geometry where the electrodes
are located in a single plane and the ions confined above this
plane. This device is realized with a relatively simple fabrication
procedure and has important implications for quantum state
manipulation and quantum information processing using large numbers
of ions.  We confine laser-cooled $^{24}$Mg$^+$ ions approximately
40 $\mu$m above planar gold electrodes.  We measure the ions'
motional frequencies and compare them to simulations. From
measurements of the escape time of ions from the trap, we also
determine a heating rate of approximately five motional quanta per
millisecond for a trap frequency of 5.3 MHz.
\end{abstract}

\maketitle

Recent interest in coherent quantum state control and methods to
realize practical quantum information processing (QIP) has led to
impressive developments in quantum processing using several
different physical systems~\cite{arda_roadmap}. Single quantum bit
(qubit) rotations, two-qubit gates, and simple quantum algorithms
have been implemented. However, perhaps the most significant
challenge for any possible physical implementation of a quantum
processor is to devise methods that scale to very large numbers of
quantum information carriers.

The system of trapped ions is an interesting candidate for QIP
because the basic requirements~\cite{di} have been demonstrated in
separate experiments~\cite{arda_roadmap}, and several schemes for
scaling this system to large numbers of qubits have been
proposed~\cite{arda_roadmap,bible,devoe,cz2000,ki,duan04}. One
approach is based on a network of interconnected processing and
memory zones where ion qubits are selectively shuttled between
zones~\cite{bible,ki}. Within this approach, miniature linear trap
arrays~\cite{rowe,barrett,GaAs,mini} and a three layer T-junction
trap~\cite{mini} have been demonstrated. Since the speed of most
multi-ion qubit gates is proportional to the ions' motional
frequencies and these frequencies are inversely proportional to the
square of the dimensions, we would like to decrease the size of
these dimensions. To do this robustly, microfabrication techniques
are required. Three-dimensional traps have been demonstrated with
boron-doped silicon~\cite{icols} and monolithically fabricated
gallium-arsenide electrodes~\cite{GaAs}. A significant
simplification in fabrication could be achieved if all trap
electrodes reside on a single surface and the ions are trapped above
this surface~\cite{jc}.  In this case, the trapping electric fields
would be the analog of magnetic fields used in ``chip'' traps for
neutral atoms (see Ref.~\cite{folman}~and references therein).
Surface-electrode ion traps have the potential added benefit for
scaling that micro-electronics for electrode potential control can
be fabricated below the plane of the electrodes~\cite{kim}.

Recently, macroscopic charged particles have been confined in a
surface-electrode trap~\cite{chuang}. Storage of atomic ions,
however, requires substantially different experimental parameters.
In this letter we report the first demonstration of stable
confinement of atomic ions in a surface-electrode trap. The trap is
constructed with standard and scalable microfabrication processes.
We load $^{24}$Mg$^+$ into this trap, measure the motional
frequencies of the ions, and find reasonable agreement with those
determined from simulations. We also approximately determine a
motional heating rate of the ion(s) that is low enough to allow for
high fidelity logic operations.

\begin{figure}[t]
\centering\includegraphics[width=8.6cm]{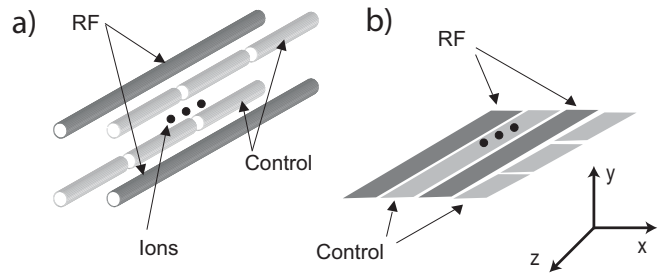}
\caption{\label{paul} (a) Standard linear RF Paul trap; (b)
Surface-electrode geometry where all electrodes reside in a single
plane, with the ions trapped above this plane.}
\end{figure}

\begin{figure}[t]
\centering \includegraphics[width=8.6cm]{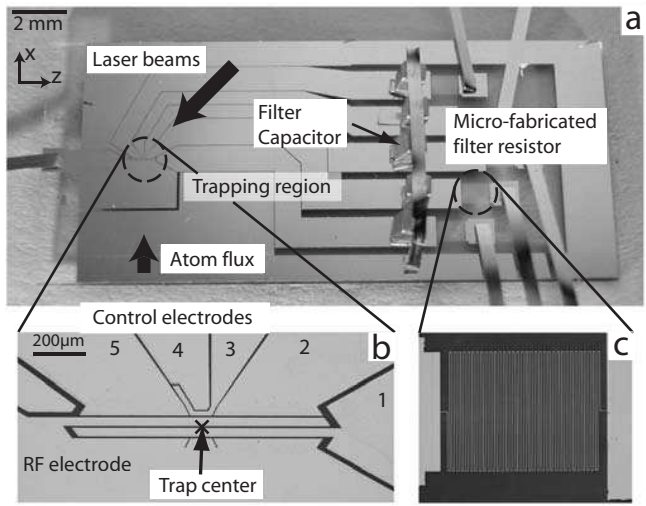}
\caption{\label{over} Pictures of the surface-electrode trap. (a)
The complete trap structure, including out-lead wires (ribbons) and
filter capacitors. The directions of the laser beams (cooling and
photoionization) and atom flux are indicated. (b) Expanded view of
the trap region (center marked by $\times$). The control electrodes
are numbered for reference in the text. (c) On-board meander line
resistor.}
\end{figure}

The standard linear radio-frequency (RF) Paul trap~\cite{paul_rev}
consists of four parallel rods whose centers are located on the
vertices of a square (Fig.~\ref{paul}a). An RF potential is applied
to two opposing rods with the other two (control electrode) rods
held at RF ground. This configuration creates a nearly harmonic
ponderomotive pseudopotential in the $\hat{x}$-$\hat{y}$ plane.
Longitudinal confinement for a single trapping zone is obtained by
segmenting each control electrode along its length and applying
appropriate static potentials to the different segments. Several
variations on this design have been
demonstrated~\cite{icols,rowe,barrett, GaAs, mini}, but it is very
desirable to simplify their construction.
A straightforward way to modify the 3-D design of Fig.~\ref{paul}a
is to place the four rods in a common plane, with alternating RF and
control electrodes~\cite{jc}; one version of this geometry is shown
in Fig.~\ref{paul}b. In this design, the rods are replaced with flat
electrodes, as shown in Fig.~\ref{over}.

We can fabricate this electrode structure by means of
photolithography and metal deposition using evaporation and
electrodeposition. For the substrate we use polished fused quartz, a
material with low RF loss.  A 0.030 $\mu$m titanium adhesion layer
and a 0.100 $\mu$m copper seed layer are first evaporatively
deposited onto the substrate.  This deposition is uniform except for
small areas for resistors where the quartz is left exposed.
Resistors ($\sim$1 k$\Omega$) and leads are fabricated through a
liftoff process that entails patterning them with standard
photolithography and evaporation of a 0.013 $\mu$m titanium adhesion
layer followed by 0.300 $\mu$m of gold. Resistors are fabricated
directly on the quartz substrate; leads are fabricated on top of the
copper seed layer. The gold electrodes near the trapping region are
electroplated onto the copper seed layer after a second
photolithographic patterning step. Afterward, the exposed initial
seed and adhesion layers are etched away to isolate electrodes and
leads. The trap electrodes are plated to a thickness of $\sim$ 6
$\mu$m so that the ratio of height to inter-electrode spacing is
relatively high (the inter-electrode spacing is $\sim$ 8 $\mu$m).
This should reduce alteration of the trapping potential due to stray
charges that may collect on the exposed insulator between
electrodes.

We create ions in the trap by photoionizing thermally evaporated
neutral magnesium atoms. The magnesium source is realized by
resistively heating a stainless steel tube containing solid
magnesium, which is sealed at both ends and has a small slit from
which evaporated magnesium atoms emerge. With a planar electrode
geometry, there is a risk of shorting electrodes to each other due
to magnesium deposited onto the trap structure.  To reduce this
risk, in a last processing step we perform a controlled hydrofluoric
acid (HF) etch of the central trap region. The HF etches away a
small part of the titanium adhesion layer and the substrate, without
affecting the electrodes. The result is a $\sim 2~\mu$m horizontal
undercut of the electrodes to help prevent shorting due to
deposition from the magnesium source. As a further precaution, we
direct the magnesium flux nearly parallel to the surface and avoid
as much as possible having the channels between electrodes be
parallel to the flux (Fig.~\ref{over}).

We use five independent control electrodes to provide sufficient
degrees of freedom to be able to overlap the electric field null
point of the static potential and the RF pseudopotential minimum. We
create a low impedance path for the RF to ground on the control
electrodes with capacitors (820 pF) that are surface mounted
directly onto the chip (in the future, capacitors could be included
as part of the fabrication process). Gold ribbons for applying the
electrode potentials are gap-welded to contact pads.


The trap structure is mounted in a copper tube that also serves as
part of an RF transformer~\cite{jefferts} and the entire structure
is surrounded by a quartz envelope. The system is baked under vacuum
prior to operation to reach a base pressure below $10^{-8}$ Pa with
the use of an ion getter pump combined with a titanium sublimation
pump.

As we describe below, the trap well depth $U_T$ for a
surface-electrode trap is fairly shallow~\cite{jc}, not much above
the mean kinetic energy of the neutral atoms before they are
ionized. Nevertheless, we can load $^{24}$Mg$^+$ ions efficiently by
resonant two-photon photoionization (PI) at $285$ nm~\cite{drewsen}.
The PI laser, resonant with the 3s$^{2~1}$S$_0$ $\leftrightarrow$
3s3p~$^{1}$P$_1$ electric dipole transition in neutral magnesium,
co-propagates with a Doppler cooling beam tuned approximately 400
MHz below the 3s~$^{2}$S$_{1/2}$ $\leftrightarrow$
3p~$^{2}$P$_{1/2}$ electric dipole transition in $^{24}$Mg$^+$ at
$280$ nm. The laser beams are parallel to the trap surface, and at
an angle of approximately 45 degrees with respect to the trap
$\hat{z}$ axis as shown in Fig.~\ref{over}a.

Since the laser beam direction has significant overlap with all
principal trap axes, cooling will be efficient in all
directions~\cite{jc}. During loading, both the Doppler cooling and
PI beams have 2 mW power and waists of $\sim$ 40 $\mu$m. The atomic
flux of magnesium intersects the laser beams at the trap
(Fig.~\ref{over}a). The cooling beam is applied continuously, while
the PI beam needs to be applied for only a few seconds to create
ions in the trap. Ions are detected by observing 3p~$^{2}$P$_{1/2}~
\rightarrow$ 3s~$^{2}$S$_{1/2}$ fluorescence along a direction
perpendicular to the trap surface with a CCD camera as in the view
of Fig.~\ref{over}b. Despite the fact that the center of the laser
beam is only 40 $\mu$m above the surface, which increases the risk
of scatter from light striking the trap electrodes, the
signal-to-background ratio for scattered light from the ions is
greater than 100 when the Doppler cooling laser is tuned
approximately 20 MHz (one-half linewidth) below resonance with
intensity slightly below saturation.

We measure the oscillation frequencies for a single ion in the trap
(equal to the center-of-mass mode frequencies for multiple ions) by
applying an oscillating field to a control electrode and observing a
change in fluorescence rate when the frequency of the applied field
is equal to one of the motional frequencies, thereby heating the
ion~\cite{jefferts}. To excite the axial mode, we apply the
oscillating field to electrode 2, while both transverse modes can be
excited using electrode 1.

As an aid in initially determining the correct operating conditions,
trapping potentials are determined using numerical solvers (boundary
element method) subject to the constraint that the RF
pseudopotential minimum overlaps the null points of the electric
field from the static potential, to minimize RF micromotion (see for
example~\cite{berkeland}). For the experiments described here, the
static potentials on each control electrode, expressed as a fraction
of the potential $V_5$ on electrode 5 (Fig.~\ref{over}b) are $V_1 =$
0.320, $V_2 =$ 0.718, $V_3 =$ 0.738, and $V_4 =$ -0.898.

\begin{table}[b]    
\caption{\label{freq} Oscillation frequencies (experimental
measurements (Exp.) and simulated values (Sim.)) for three different
potential configurations explained in the text. The axial frequency
is denoted $f_{\parallel}$, while $f_{\perp 1}$ and $f_{\perp 2}$
are the frequencies of the two transverse modes whose axes are
indicated by the cross in Fig.~\ref{pot}. The uncertainties in the
experimental values for the frequencies are approximately 0.10 MHz.}
\begin{tabular}{rr | rrrr l}\\
$V_5~$ & $V_{\rm RF}~$ & $f_{\parallel}~~$ & $f_{\perp 1}~~$ & $f_{\perp 2}~~$ & $U_T~~$  \\
(V)   &(V)    & ~~(MHz)           & (MHz)         & (MHz)         & (meV) \\
\hline
\multirow{2}{*}{5.00} & \multirow{2}{*}{103.5}
 & 2.83 & 15.78 & 17.13 &       &~~~(Exp.)\\
&& 2.77 & 15.67 & 17.21 & ~~177$\pm$ 6 &~~~(Sim.)\\
\hline
\multirow{2}{*}{2.00} & \multirow{2}{*}{103.5}
&  1.84 & 15.87 & 16.93&        &~~~(Exp.)\\
&& 1.75 & 16.23 & 16.87& ~~193$\pm$ 11 &~~~(Sim.)\\
\hline
\multirow{2}{*}{5.00} & \multirow{2}{*}{46.2}
&  2.85 & 5.28 & 8.29&          &~~~(Exp.)\\
&& 2.77 & 5.05 & 8.75&    ~~~6$\pm$ 1 &~~~(Sim.)\\
\end{tabular}
\end{table}

\begin{figure}[t]\fbox{
\centering\includegraphics[angle=0,width=8.6cm]{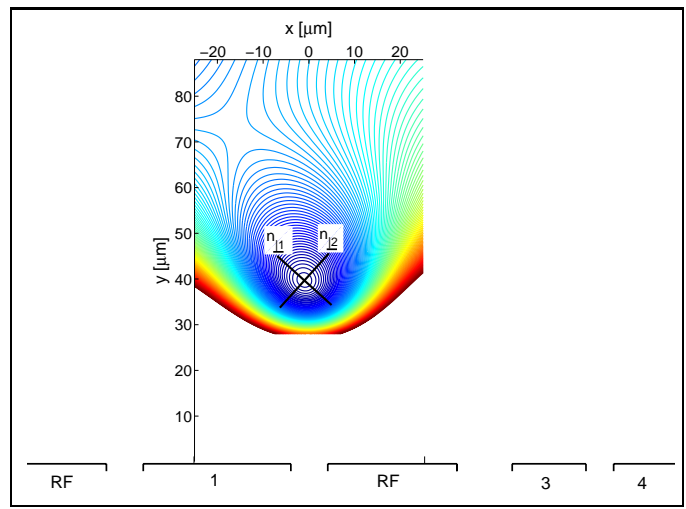}}
\caption{\label{pot} A transverse cross-section of the simulated
trapping potential (both RF pseudopotential and static potentials
included) for potentials corresponding to $V_5=5.00$ V and $V_{\rm
RF} =$ 103.5 V ($U_T =$ 177 meV). The cross indicates the directions
of the normal mode axes $\mathrm{n}_{\perp 1}$ and
$\mathrm{n}_{\perp 2}$, and the expected position for an ion at the
center of the trap in the $\hat{x}$-$\hat{y}$ plane.  The separation
between contour lines corresponds to 5 meV and the colors blue to
red correspond to low to high potential. The electrodes are depicted
to scale in the lower part of the figure and labeled as in
Fig.~\ref{over}b.}
\end{figure}

The peak potential amplitude $V_{\rm RF}$ applied to the RF
electrode (at a frequency of 87 MHz) is difficult to measure
directly.  To determine it, we measure the three mode frequencies
for a fixed RF power and $V_5 =$ 5.00 V.  We use the simulations
with the measured electrode dimensions to extract a best fit value
of $V_{\rm RF} =$ 103.5 V. The consistency between the experimental
and simulated frequencies for this fit is shown in Table~\ref{freq}.
We further check for consistency by comparing the measured and
predicted mode frequencies for two other values of the control and
RF potentials (also shown in Table~\ref{freq}). The value $V_{\rm
RF}=46.2$ V is determined by scaling the RF power delivered to the
trap. The agreement between predicted and measured frequencies is
within a few percent. Finally, we determine numerically the overall
potential depth $U_T$ in the pseudopotential
approximation~\cite{pseud1,pseud2} for the different control and RF
potentials (also shown in Table~\ref{freq}). A transverse
cross-section (the $\hat{x}$-$\hat{y}$ plane of Fig.~\ref{paul}b) of
the trapping potential for $V_5=5$ V and $V_{\rm RF}=103.5$ V near
the central trap region is shown in Fig.~\ref{pot}.

\begin{figure}
\centering \includegraphics[width=8.6cm]{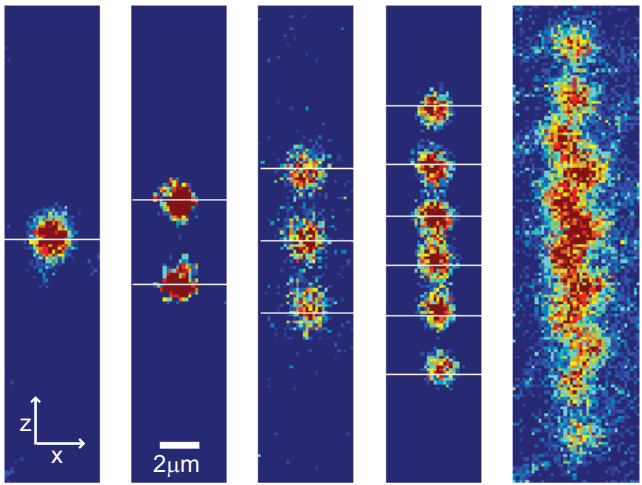}
\caption{\label{ioner} False-color images of one, two, three, six,
and 12 ions loaded into the surface trap (red corresponds to highest
fluorescence count rate). The length scale is determined from a
separate image of the electrodes whose dimensions are known. The
horizontal bars indicate the separation distance between the ions as
predicted from the measured axial oscillation frequency. The ratio
between transverse and axial oscillation frequencies makes it
energetically favorable for the 12 ion string to break into a
zig-zag shape (see for example~\cite{dubin}).}
\end{figure}


In Fig.~\ref{ioner}, we show groups of ions for the $V_{\rm RF} =
103.5$ V and $V_5 = 2.00$ V configuration. The separation of the
ions is related to the center-of-mass axial oscillation
frequency~\cite{bible}, and the horizontal bars indicate the
expected ion separations according to the measurement of this
frequency (1.84 $\pm$ 0.10 MHz for this configuration). When the
number of ions becomes large enough, the string breaks into a
zig-zag configuration (see for example~\cite{dubin}).

Because $^{24}$Mg$^+$ lacks hyperfine structure, we cannot easily
determine heating rates near the quantum limit (without applying
large magnetic fields) due to poor internal state
discrimination~\cite{monroe_cooling}. However, an approximate value
of the transverse mode heating rate can be determined by observing
how long an ion remains trapped in a shallow well in the absence of
the Doppler cooling light~\cite{GaAs}. If we assume the heating rate
(the energy change per unit of time) is constant until the ion can
overcome the saddle-point in Fig.~\ref{pot}, we can estimate the
initial heating rate as $\langle r \rangle \simeq U_T/(\langle
\tau_s \rangle h f_{\perp 1})$ quanta per second, where $\langle
\tau_s \rangle$ is the average survival time of the ion in absence
of cooling light, $h$ is Planck's constant and $f_{\perp 1}$ is the
smaller of the two transverse mode frequencies.


To help ensure that the results are not influenced by the ions
thermalizing to the ambient temperature ($\sim$ 300 K $\simeq$ 25
meV) we perform the measurements of $\langle \tau_s \rangle$ with a
very shallow well depth, $U_T = 6~ \pm$ 1 meV ($V_5=5$ V, $V_{\rm
RF} =$ 46.2 V, and $f_{\perp 1} \simeq$ 5.3 MHz). From these data,
we determine $\langle \tau_s \rangle =53\pm 10$ s, and find $\langle
r \rangle \simeq$ 5 quanta per millisecond.


Johnson noise in the resistance of the RC filters on the control
electrodes is a potential source of heating~\cite{bible}, but we
theoretically estimate it to contribute only 1 quantum per second.
Therefore the observed heating is apparently dominated by anomalous
heating as observed in previous experiments~\cite{turchette}.  We
note that in presence of Doppler cooling light, the lifetime of the
ion is several hours (in the trap with $U_T =$ 177 meV), presumably
limited by chemical reactions~\cite{molhave}.

Heating and decoherence rates for small ion traps will be an
important issue for future implementations of quantum information
processing, and will need to be more thoroughly investigated.
Therefore, an important next step is to replace the $^{24}$Mg$^+$
with an isotope that allows for Raman sideband cooling, such as
$^{25}$Mg$^+$ or $^{9}$Be$^+$, and to study heating rates near the
ground state of motion~\cite{turchette,monroe_cooling}. A
longer-term goal is to design and fabricate a surface trap with more
complex structures involving T- or X-junctions with separate memory-
and processing zones.


We thank E. Langlois and J. Moreland for advice on the
micro-fabrication processes and Y. Le Coq for helpful comments on
the manuscript. This work was supported by the Advanced Research and
Development Activity (ARDA) under contract MOD-7171.05 and NIST. S.
S. wishes to thank the Carlsberg Foundation for financial support.
This manuscript is a publication of NIST and not subject to U. S.
copyright.

\end{document}